\PassOptionsToPackage{prologue,dvipsnames}{xcolor}

\documentclass[sigconf]{acmart}

\AtBeginDocument{%
  }

\setcopyright{none}
\renewcommand\footnotetextcopyrightpermission[1]{} 

\usepackage{multirow}
\usepackage{graphicx}
\usepackage{diagbox}
\usepackage{subcaption}
\usepackage{xspace}
\usepackage[ruled, vlined, linesnumbered]{algorithm2e}
\usepackage{bbding}
\usepackage{threeparttable}
\usepackage{amsmath}
\usepackage{adjustbox}
\usepackage{listings}
\usepackage[T1]{fontenc}
\usepackage{semantic}
\usepackage{newfloat}
\usepackage{enumitem}
\usepackage{booktabs}
\usepackage{arydshln}
\usepackage{makecell}

\usepackage{framed}
\usepackage{hyperref}
\hypersetup{hidelinks,
	colorlinks=true,
	allcolors=black,
	pdfstartview=Fit,
	breaklinks=true}

\usepackage{pifont}
\newcommand{\CodeIn}[1]{{\small\texttt{#1}}}

\DeclareFloatingEnvironment[fileext=frm,placement={!ht},name=Rules]{infrule}

\lstdefinestyle{base}{
  language=java,
  moredelim=**[is][\color{ForestGreen}]{~}{~},
  moredelim=**[is][\color{BrickRed}]{!}{!},
}

\lstdefinestyle{desc}{
    backgroundcolor = \color{yellow!10},    
    numbers = none,                     
}

\newcommand\detector{CompVPD}

\newcommand{\detectorContext}[1]{$\mbox{CompVPD}^{#1}$}
\newcommand{\detectorWindowContext}[2]{$\mbox{CompVPD}_{#1}^{#2}$}

\newcommand\dataset{VulFix}
\newcommand\vulfixminer{VulFixMiner}
\newcommand\colefunda{CoLeFunDa}
\newcommand\midas{MiDas}
\newcommand\sun{Sun et. al.}
\newcommand\deepjit{DeepJIT}

\definecolor{deep-blue}{RGB}{0,0,200}
\definecolor{deep-red}{RGB}{200,0,0}

\begin{document}

\title{CompVPD: Iteratively Identifying Vulnerability Patches Based on Human Validation Results with a Precise Context}


\author{Tianyu Chen}
\email{tychen811@pku.edu.cn}
\affiliation{%
  \institution{Key Lab of HCST (PKU), MOE; SCS; Peking University}
  \city{Beijing}
  \country{China}
}

\author{Lin Li}
\email{lilin88@huawei.com}
\affiliation{%
  \institution{Huawei Cloud Computing Technologies Co., Ltd.}
  \city{Beijing}
  \country{China}
}

\author{Taotao Qian}
\email{qiantaotao@huawei.com}
\affiliation{%
  \institution{Huawei Cloud Computing Technologies Co., Ltd.}
  \city{Beijing}
  \country{China}
}

\author{Jingyi Liu}
\email{liujingyi7@huawei.com}
\affiliation{%
  \institution{Huawei Cloud Computing Technologies Co., Ltd.}
  \city{Beijing}
  \country{China}
}

\author{Wei Yang}
\email{Wei.Yang@UTDallas.edu}
\affiliation{%
  \institution{University of Texas at Dallas}
  \city{Dallas}
  \country{USA}
}

\author{Ding Li}
\email{ding_li@pku.edu.cn}
\affiliation{%
  \institution{Key Lab of HCST (PKU), MOE; SCS; Peking University}
  \city{Beijing}
  \country{China}
}

\author{Guangtai Liang}
\email{liangguangtai@huawei.com}
\affiliation{%
  \institution{Huawei Cloud Computing Technologies Co., Ltd.}
  \city{Beijing}
  \country{China}
}

\author{Qianxiang Wang}
\email{wangqianxiang@huawei.com}
\affiliation{%
  \institution{Huawei Cloud Computing Technologies Co., Ltd.}
  \city{Beijing}
  \country{China}
}

\author{Tao Xie}
\email{taoxie@pku.edu.cn}
\affiliation{%
  \institution{Key Lab of HCST (PKU), MOE; SCS; Peking University}
  \city{Beijing}
  \country{China}
}

\renewcommand{\shortauthors}{Anonymous}

\begin{abstract}
Applying security patches in open source software timely is critical for ensuring the security of downstream applications. However, it is challenging to apply these patches promptly because notifications of patches are often incomplete and delayed. To address this issue, existing approaches employ deep-learning (DL) models to identify additional vulnerability patches by determining whether a code commit addresses a vulnerability. Nonetheless, these approaches suffer from low accuracy due to the imprecise context provided for the patches.
To provide precise context for patches, we propose a multi-granularity slicing algorithm and an adaptive-expanding algorithm to accurately identify code related to the patches. 
Additionally, the precise context enables to design an iterative identification framework, \detector{}, which utilizes the human validation results, and substantially improve the effectiveness.
We empirically compare \detector{} with five state-of-the-art/practice (SOTA) approaches in identifying vulnerability patches. The results demonstrate that \detector{} improves the F1 score by 20\% compared to the best scores of the SOTA approaches.
Additionally, \detector{} contributes to security practice by helping identify 20 vulnerability patches and 18 fixes for high-risk bugs from 2,500 recent code commits in five highly popular open-source projects.

\end{abstract}




\maketitle
\section{Introduction}~\label{sec: intro}

Open source software (OSS) is critical and widely used in software development~\cite{wang2020empirical}. However, OSS also harbors numerous vulnerabilities, which threaten the security of downstream applications. One method to secure OSS usage is the swift adoption of security patches that address known vulnerabilities. Nevertheless, applying security patches promptly is particularly challenging due to delayed and incomplete knowledge about these patches. For instance, a recent study~\cite{li2017large} indicates that it may take more than a year for the National Vulnerability Database (NVD)~\cite{nvd}, the most well-known vulnerability knowledge base, to recognize a security patch. Furthermore, OSS maintainers may release vulnerability patches silently due to inadequate patch management~\cite{wang2019detecting}, complicating efforts to monitor these patches effectively.

To assist developers in swiftly adopting security patches, researchers have proposed numerous approaches to automatically identify security patches~\cite{vulfixminer, colefunda, zhou2021spi, espi, vulcurator, deepjit, midas, graphspd}. The high level idea behind these methods is to train a machine learning model capable of predicting whether a patch addresses a vulnerability. However, these existing approaches generally suffer from limitations such as low accuracy. For instance, one of the more recent methods, MiDas~\cite{midas}, achieves an F1 score of only 0.200.

One of the reasons for the low accuracy of existing approaches is their failure to include precise context for patches. Some approaches~\cite{vulfixminer, deepjit, midas, sunVulPatch} only consider the changed lines and ignore the comprehensive context, such as control and data flow dependencies, of the patches. These contexts are crucial for explaining the usage of code commits. Conversely, other approaches~\cite{colefunda, zhou2021spi, espi, vulcurator, graphspd} consider the entire changed methods or files as the context of a patch, which is too coarse-grained. Furthermore, using an excessive amount of context significantly increases the data volume for the models, which in turn limits the size of the models. Our empirical results in Table~\ref{tab: dataset distribution} show that the average token number is 7992. Such a large number of tokens leads to substantial efficiency costs in fine-tuning and inference, as the efficiency and memory costs of transformer-based DL models are quadratic with context size. Therefore, these approaches employ only lightweight models, e.g., RGCN~\cite{chen2019rgcn} with manually designed embedding functions, which also contributes to low accuracy.

In this paper, we design two novel algorithms that generate the precise (i.e., reduced comprehensive) context for code commits. First, we introduce a multi-granularity slicing algorithm to remove irrelevant components (i.e., files, methods, and statements) from a code commit. Our rationale behind this algorithm is to eliminate components that hinder existing approaches from leveraging comprehensive contexts effectively. We exclude irrelevant components, specifically those lacking control-flow and data-flow dependencies on changed lines. These components often constitute a large portion (e.g., more than 99\%~\cite{codechange}) and are unlikely to be included in the sequence of statements relevant to exploiting the corresponding vulnerability or its patch. Additionally, we remove the bodies of relevant methods that do not contain changed lines, retaining only their annotations~\cite{annotation} and signatures, which succinctly describe their functionalities.
Second, we devise an adaptive context-expanding algorithm to select the closest code hunks as a reduced context. Specifically, we expand the context at the block level; for each statement block in a code commit's context, we either include the entire block or exclude it. Our rationale is that statements within the same block are highly correlated, and thus, we avoid splitting them during context selection. We then prioritize the closest blocks as the most significant contexts of a given code commit, within a maximum context size.


Considering that our precise contexts significantly improve the fine-tuning and inference efficiency, we have also been able to design an iterative identification framework named \detector{}. For a set of input code commits, \detector{} iteratively fine-tunes its deep learning (DL) model using a small set of human-validated results (i.e., whether a code commit is a vulnerability patch) and then employs the updated model to further identify vulnerability patches. Our rationale is based on \detector{}'s application in industrial practice, where identified results must be manually validated by security engineers.
We do not require security engineers to validate all code commits simultaneously; instead, we use the validation results to enhance \detector{}'s effectiveness (by reducing false positives), thus diminishing the manual validation costs. Additionally, while existing approaches struggle to identify code commits from an "unseen" repository (one not present in the training set), our iterative fine-tuning framework effectively transforms an "unseen" repository into a "seen" one. This is achieved by iteratively fine-tuning on the identified code commits with manual validation, thereby addressing the generalization limitations~\cite{icse2023empirical} of existing models.



We empirically compare \detector{} with five state-of-the-art approaches for identifying Java vulnerability patches, namely \deepjit{}~\cite{deepjit}, \vulfixminer{}~\cite{vulfixminer}, \sun{}~\cite{sunVulPatch}, \midas{}~\cite{midas}, and RGCN~\cite{chen2019rgcn}, using a widely used dataset provided by \vulfixminer{} (referred to as \dataset{}). Our evaluation results reveal several findings that demonstrate \detector{}'s effectiveness and efficiency.

\detector{} improves the F1 score by 20\% compared to the best scores of these state-of-the-art approaches. Additionally, our iterative identification framework is highly effective, identifying 35\% (CodeBERT) and 13\% (StarCoder) more vulnerability patches within the same human efforts. \detector{} efficiently processes one code commit in less than 0.5 seconds. Each component of our proposed context improves \detector{}'s effectiveness by at least 3\% in terms of AUC and at least 12\% in terms of F1 score.

Furthermore, \detector{} provides significant value to security practice by helping to identify 20 vulnerability patches and 18 fixes for high-risk bugs from 2,500 recent code commits of five highly popular open-source projects, namely \CodeIn{jenkins}, \CodeIn{hutool}, \CodeIn{dubbo}, \CodeIn{hiro}, and \CodeIn{light-4j}.


In summary, this paper makes the following main contributions:
\begin{itemize}
    \item We propose two novel algorithms to generate precise contexts for code commits, enhancing the accuracy of vulnerability patch identification.
    \item We introduce an iterative identification framework that leverages human validation results of identified vulnerability patches to refine and improve the identification process.
    \item We conduct a comprehensive evaluation to demonstrate the effectiveness and efficiency of our approach. Our approach successfully identifies 20 vulnerability patches and 18 fixes for high-risk bugs in real-world practice.
\end{itemize}

\section{Motivation}\label{sec:background}

\begin{figure}
    \centering    
\begin{lstlisting}[style=base, caption = An example Vulnerability Patch of CVE-2017-4971\protect\footnotemark, label = lst: cve-2017-4971-raw]
`Commit Message: Use fixed parser for empty value binding expressions`
!--- spring-webflow/src/main/java/org/springframework/webflow/mvc/view/AbstractMvcView.java!
~+++ spring-webflow/src/main/java/org/springframework/webflow/mvc/view/AbstractMvcView.java~
@@ -14,6 +14,7 @@
 import org.springframework.binding.expression.Expression;
 import org.springframework.binding.expression.ExpressionParser;
 import org.springframework.binding.expression.ParserContext;
~+import org.springframework.binding.expression.beanwrapper.BeanWrapperExpressionParser;~
 import org.springframework.binding.expression.support.FluentParserContext;
 import org.springframework.binding.expression.support.StaticExpression;
 import org.springframework.binding.mapping.MappingResult;

@@ -48,6 +49,7 @@
   private org.springframework.web.servlet.View view;
   private RequestContext requestContext;
   private ExpressionParser expressionParser;
~+  private final ExpressionParser emptyValueExpressionParser=`\textcolor{ForestGreen}{new}` BeanWrapperExpressionParser()~
   private ConversionService conversionService;
   private Validator validator;
   private String fieldMarkerPrefix = "_";

@@ -257,7 +259,7 @@
   */
   protected void addEmptyValueMapping(DefaultMapper mapper, String field, Object model) {
     ParserContext parserContext = new FluentParserContext().evaluate(model.getClass());
!-    Expression target = expressionParser.parseExpression(field, parserContext);!
~+    Expression target = emptyValueExpressionParser.parseExpression(field, parserContext);~
     try {
       Class<?> propertyType = target.getValueType(model);
       Expression source = new StaticExpression(getEmptyValue(propertyType));
...//10 more lines in this method, 295 more lines in this file
\end{lstlisting}
\end{figure}

\footnotetext{\url{https://github.com/spring-projects/spring-webflow/commit/57f2ccb66946943fbf3b3f2165eac1c8eb6b1523}}

A vulnerability patch is defined as one code commit that addresses a security vulnerability of a code repository.
The target of identifying vulnerability patches is to determine whether a given code commit fixes a vulnerability.
A code commit consists of a commit message, a set of deleted lines (i.e., lines in red) and added lines (i.e., lines in green).
For example, Listing~\ref{lst: cve-2017-4971-raw} shows the vulnerability patch of CVE-2017-4971~\footnote{\url{https://nvd.nist.gov/vuln/detail/CVE-2017-4971}}.
It fixes an Insecure-Default-Initialization-of-Resource (CWE-1188) vulnerability by changing the expression parser in Line 26 to a specific empty-expression parser in Line 27.

\subsection{Limitations of Existing Approaches}
Existing approaches~\cite{vulfixminer, colefunda, zhou2021spi, espi, vulcurator, deepjit, midas, graphspd} suffer from low accuracy mainly because they fail to include the precise context for input code commits.

\subsubsection{Insufficient Context}
Existing sequential-based approaches~\cite{vulfixminer, deepjit, midas, sunVulPatch} consider only the changed lines of a code commit and ignores its comprehensive context, such as the commit's control and data flow dependencies. 
However, these contexts help explain the usage of code commits. 
For example, in Listing~\ref{lst: cve-2017-4971-raw}, these approaches consider only Line 8, 17, 26, and 27, overlooking the contexts after Line 27, which explains the usage of this vulnerability patch, i.e., using the specific empty-expression parser defined in Line 27 to ensure that the input String \CodeIn{field} does not contain malicious expressions.

\subsubsection{Too-Coarse Context}
Existing graph-based approaches~\cite{colefunda, zhou2021spi, espi, vulcurator, graphspd} consider the whole changed methods/files as the context of a patch, which is too coarse grained.  
Listing~\ref{lst: cve-2017-4971-raw} also shows that after Line 30, there are 10 more lines in its corresponding method, \CodeIn{addEmptyValueMapping}, and 295 more lines in its corresponding file, \CodeIn{AbstractMvcView.java}.
Considering all these contexts substantially increases the data volume for the models, which further limit the size of the models. 
Our empirical results in Table~\ref{tab: dataset distribution} shows that their average token number is 7992.
Such a large number of token numbers leads to a substantial efficiency costs in fine-tuning/inference as the efficiency and memory costs of a transformer-based DL models are quadratic with the context size. 

\subsection{Our insight of a Precise Context}
To address the limitations faced by existing approaches, we need to trade-off the context size of a code commit with the context size of a DL model/LLM, i.e., generating a code commit's precise context.
In this subsection, we first provide a specific definition of a code commit's comprehensive contexts and explain our high-level idea of generating a precise context based on a comprehensive one.

\subsubsection{A Code Commit's Comprehensive Context}
A comprehensive context, which is defined as code snippets with control/data flow dependencies with a code commit's changed lines, helps explain the functionalities and usages of a code commit and thus provide evidence to comprehend code commits and identify vulnerability patches.
We use the example in Listing~\ref{lst: cve-2017-4971-reduce} to show how a comprehensive context (including both control/data-flow contexts and method-invocation contexts) helps identify vulnerability patches.
Now that this code commit changes the definition of a variable \CodeIn{target}, whether it fixes a vulnerability depends on whether the original definition of \CodeIn{target} is vulnerable, i.e., can be exploited by attackers.
Control/data-flow contexts include the usage of \CodeIn{target} in Lines 23-29.
These lines indicate that \CodeIn{target} is added into a global data structure \CodeIn{mapper} without validation.
Thus, \CodeIn{target} with malicious expression can lead to potential vulnerabilities. 

\begin{figure}
    \centering
 \begin{lstlisting}[style=base, caption = A Precise Context of CVE-2017-4971's Patch, label = lst: cve-2017-4971-reduce]
`Commit Message: Use fixed parser for empty value binding expressions`
!--- spring-webflow/src/main/java/org/springframework/webflow/mvc/view/AbstractMvcView.java!
~+++ spring-webflow/src/main/java/org/springframework/webflow/mvc/view/AbstractMvcView.java~
@@ -2,1 +2,2 @@
~+import org.springframework.binding.expression.beanwrapper.BeanWrapperExpressionParser;~
 public abstract class AbstractMvcView implements View {
@@ -8,1 +9,2 @@
   private ExpressionParser expressionParser;
~+  private final ExpressionParser emptyValueExpressionParser=`\textcolor{ForestGreen}{new}` BeanWrapperExpressionParser()~
@@ -44,7 +46,7 @@
/* Adds a special {@link DefaultMapping} that results in setting the target field */
   protected void addEmptyValueMapping(DefaultMapper mapper, String field, Object model){
     ParserContext parserContext=new FluentParserContext().evaluate(model.getClass());
!-    Expression target = expressionParser.parseExpression(field, parserContext);!
~+    Expression target = emptyValueExpressionParser.parseExpression(field, parserContext);~
     try {
       Class<?> propertyType=target.getValueType(model);
       Expression source = new StaticExpression(getEmptyValue(propertyType));
       DefaultMapping mapping=new DefaultMapping(source,target);
       mapper.addMapping(mapping);
     }
\end{lstlisting}
\vspace{-0.3cm}
\end{figure}

\subsubsection{High-Level of Our Proposed Approach}
Our proposed approach involves a two-step process to generate a comprehensive context within a limited context size.
Our first step is removing a code commit's irrelevant components (i.e., files, methods, and statements).
Thus all the rest components correspond to its comprehensive context.
We drop irrelevant components, i.e., those without control-flow and data-flow dependencies on changed lines, as they account for a large portion (e.g., more than 99\%~\cite{codechange}) and will likely not be included in the sequence of statements relevant to exploiting the corresponding vulnerability or its patch to prevent the exploit. 
We also remove the bodies of relevant methods without changed lines and keep only their annotations~\cite{annotation} and signatures, which have already described their functionalities in a few words.
Second, we devise an adaptive context-expanding algorithm to select the closest code hunks as a reduced context. 
Specifically, we expand the context at the block level; for each statement block in a code commit's context, we either include the entire block or exclude it. 
For example, we include Lines 16-21 (the \CodeIn{try} block) in Listing~\ref{lst: cve-2017-4971-reduce} and remove its following statement blocks, e.g., a \CodeIn{catch} block.
Our rationale is that statements within the same block are highly correlated, and thus, we avoid splitting them during context selection. 
We then prioritize the closest blocks as the most significant contexts of a given code commit, within a maximum context size.


\subsection{Iterative Identification with Human Validation Results}

Considering that our precise contexts significantly improve the fine-tuning and inference efficiency, we further design an iterative identification framework. 
Our rationale comes from \detector{}'s usage in industrial practice, i.e., the identified results need to be manually validated by security engineers to determine whether they correspond to a vulnerability patch.
With the help of validation results, we can convert a cross-project scenario (identifying commits from a repository not included in the training set), which is common in industrial practice, into a mixed-project scenario (identifying commits from a repository included in the training set), which substantially improves the accuracy of \detector{} (and existing approaches).
For a set of code commits in an ``unseen'' repository, \detector{} iteratively fine-tunes its DL model with the ground-truths of its identified results and then identifies new vulnerability patches.

\section{Approach}~\label{sec: approach}

\begin{figure}[t]
\centering
\includegraphics[width=1\linewidth]{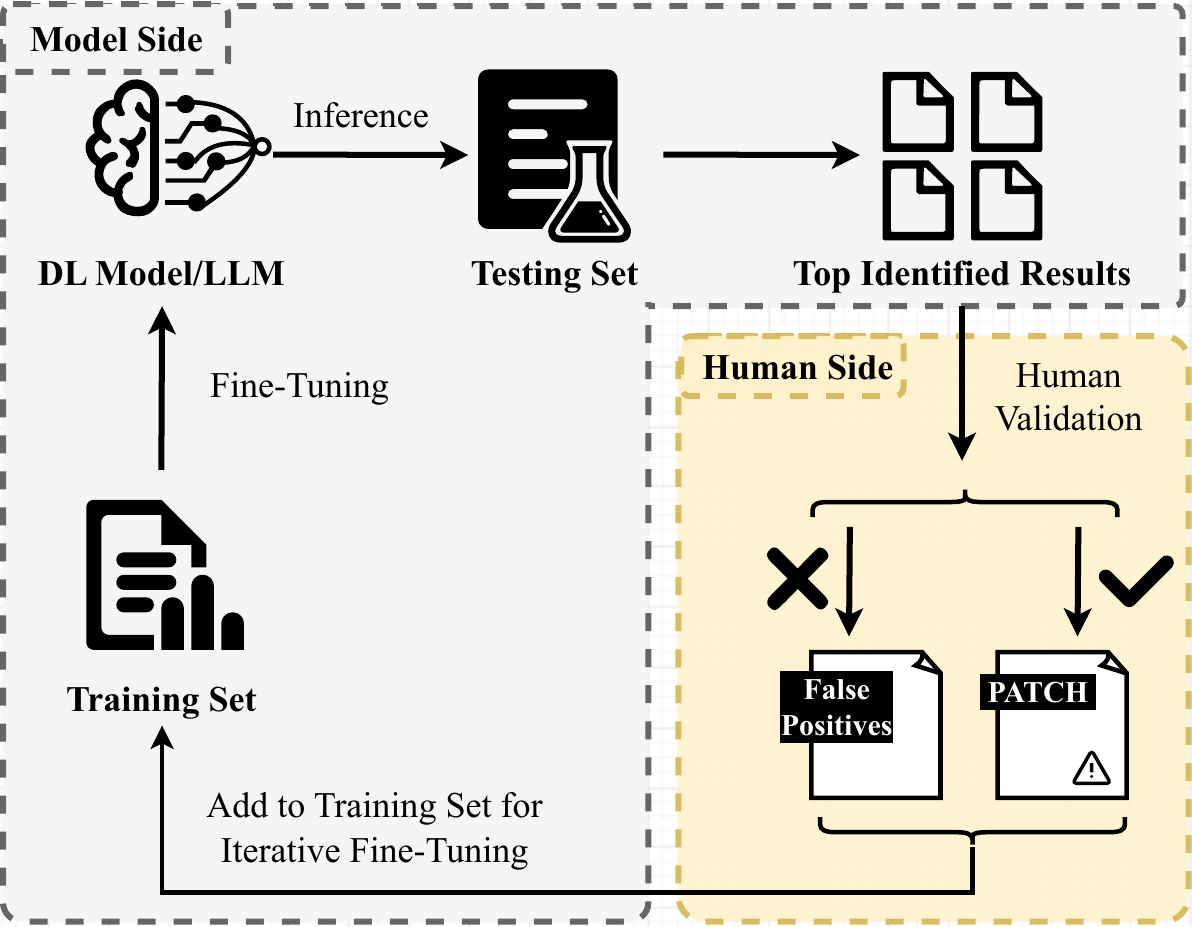}
\caption{High-level framework of \detector{}}
\label{fig: framework}
\end{figure}

\begin{figure*}[t]
\centering
\includegraphics[width=1\linewidth]{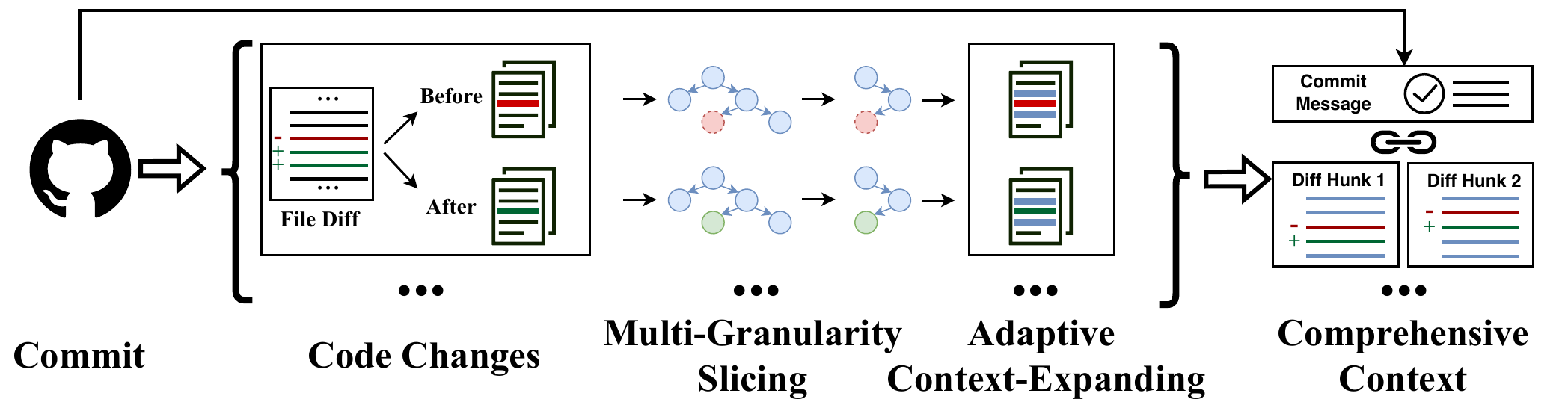}
\caption{\detector{}'s Workflow in Generating a Precise Context}
\label{fig: context workflow}
\end{figure*}

In this section, we propose \detector{}, the first approach to iteratively identify vulnerability patches by utilizing the human validation results of identified vulnerability patches.
Additionally, our approach leverages a code commit's precise context to further help LLMs comprehend the definition, usage, and invocation sequence of code changes, thus introducing higher identification accuracy.

The workflow of \detector{} is illustrated in Figure~\ref{fig: framework}.
\detector{} takes a series of code commits as input and outputs each commit's probability of whether it fixes a vulnerability.
For each code commit, \detector{} generates its comprehensive context for both fine-tuning and inference.
Then, \detector{} iteratively works with the following three steps.
First, \detector{} is fine-tuned on a given training set that consists of labeled vulnerability patches and commits with other functionalities.
Second, \detector{} identifies vulnerability patches among the given set of code commits.
Third, our security engineers manually validate whether these code commits are vulnerability patches, and add these validated results to the training set for another round of fine-tuning.



\subsection{Context Generation}
To balance the context size with the training costs of Large Lange Models (LLMs), we propose two novel algorithms, multi-granularity slicing and adaptive context-expanding, to generate a code commit's precise (reduced comprehensive) context within a limited window size.
The input of this step is a given code commit and it outputs a commit file with a precise context comprehended by LLMs.
Specifically, this commit file includes the commit message of a given code commit and a series of code changes with precise contexts.

In context generation, \detector{} works in the following four steps.
First, it extracts source codes before and after a given code commit.
Second, it conducts multi-granularity slicing over the abstract syntax trees (ASTs) of these files, thus removing irrelevant components of these files.
Third, it conducts adaptive context-expanding over these reduced files.
Fourth, it concatenates the commit messages with the output of step 3 as the input of our following model.

\subsubsection{Multi-Granularity Slicing}
To remove a commit’s irrelevant contexts, we design our multi-granularity slicing algorithm.
Our target of this step is to remove components (i.e., files, methods, and statements) that prevent existing methods from leveraging comprehensive contexts.
The input of this algorithm is all files in an open-source repository with a set of changed lines.
The output of this algorithm is a series of reduced files that include the relevant codes of all code changes.

At the file level, we remove all non-Java files and test files as we notice that non-Java files and test files do not correspond to the functionalities of a Java repository.
At the method level, we remove methods without invocation dependencies (either direct or indirect ones) of changed methods because they will not occur in the invocation sequence of each changed method.

At the statement level, we design a Bi-BFS slicing algorithm for slicing.
Overall, we remove all statements that have no control-flow and data-flow dependencies with changed lines as they count for a large portion (e.g., more than 99\%~\cite{codechange}) and will not occur in the invocation sequence of vulnerabilities and patches.

\begin{figure*}[t]
\centering
\includegraphics[width=1\linewidth]{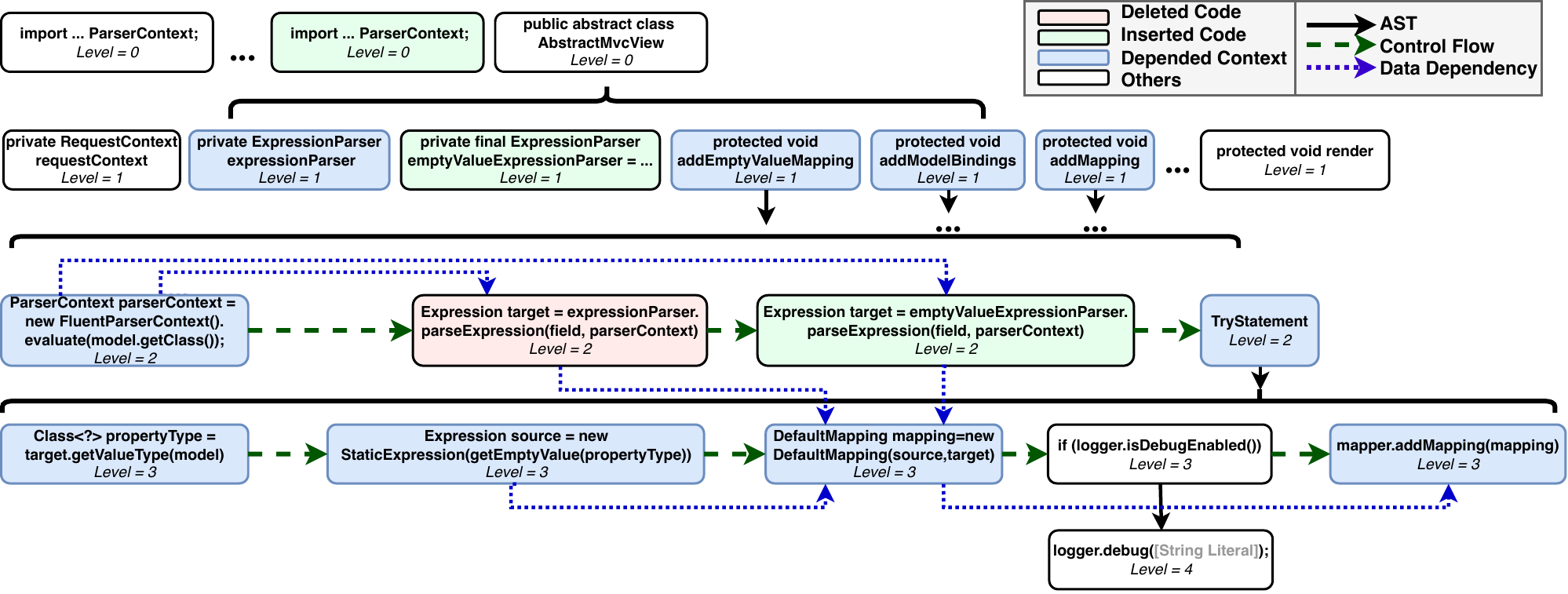}
\caption{The Code Property Graph of CVE-2017-4971}
\label{fig: cpg}
\end{figure*}

Figure~\ref{fig: cpg} shows an example of multi-granularity slicing.
It includes the AST of the changed Java file shown in Listing~\ref{lst: cve-2017-4971-raw}.
Here, we mark the deleted and inserted codes, and their context codes as red, green, and blue nodes, respectively.
In this figure, caller and callee methods, such as \CodeIn{addEmptyValueMapping}, are kept while irrelevant methods, such as \CodeIn{render}, are removed after slicing. 
Additionally, statements without control-flow and data-flow dependencies, e.g., \CodeIn{logger.debug()}, are also removed.

\subsubsection{Adaptive Context-Expanding}

To generate a precise context of a given code commit, we design our adaptive context-expanding algorithm.
Here, a ``complete'' context means that it either contains a whole statement block or does not contain this statement block.
For example, Line 22 in Listing~\ref{lst: cve-2017-4971-raw} is part of a \CodeIn{block comment}, and the \CodeIn{try} statement in Lines 27-29 is incomplete.
When compared with straightforward contexts, e.g., the closest three lines of changed lines (used in \CodeIn{git diff}~\cite{git}), a ``complete'' context helps LLMs comprehend code changes more effectively by leveraging the hierarchy information, i.e., statements in the same block are highly correlated.
Specifically, the input of this algorithm is a pair of files before and after a code commit and outputs an integrated file of code changes with precise contexts.

This algorithm consists of three main steps.
First, \detector{} uses a third-party module, \CodeIn{com.github. difflib.DiffUtils}, to generate the change hunks when comparing the files before and after a code commit.
Second, \detector{} adaptively chooses the context width of each hunk to achieve a ``complete'' context.
Third, \detector{} extends the functionality in \CodeIn{DiffUtils} to generate diff files with each hunk's corresponding context width.

\begin{algorithm}[t]
	\SetKwData{DiffUtils}{DiffUtils}\SetKwData{integer}{\textbf{int}}
	\SetKwFunction{Union}{Union}\SetKwFunction{getDepth}{minDepth}
	\SetKwInOut{Input}{input}\SetKwInOut{Output}{output}
 
	\Input{$beforeFile, afterFile$, a pair of files before and after a commit}
        \Input{$maxWidth$, the maximum length of contexts}
	\Output{$context$, a precise context}
	\BlankLine
        
        \tcp{The iteration order of context widths}
        $mid \gets maxWidth/2$\;
        $widths := [mid,mid - 1, mid + 1,\dots, 0, maxWidth]$\;

        \tcp{Obtain one code line's minimal depth in AST}
        \SetKwProg{Fn}{Function}{ :}{end}
        \Fn{\getDepth(line: \integer)}{
            $depth \gets \inf$\;
            \For{$nodes \in AST$}{
                \If{$nodes.startLine.equals(line)$}{
                    $depth = \min(depth, AST.depth(line))$
                }
            }
            \Return{$depth$}
        }

        \tcp{Adaptive context-expanding}
        $hunks \gets \DiffUtils.changes(beforeFile, afterFile)$\;
        
        \For{$i \gets 0$ \KwTo $hunks.size()$}{
            $w_{st} \gets \mathop{\arg\max}\limits_{width[i]} \{\getDepth(hunks[i].st - width[i])\}$\;
            $w_{end} \gets \mathop{\arg\max}\limits_{width[i]} \{\getDepth(hunks[i].end + width[i])\}$\;
            $newHunks[i] \gets \DiffUtils.hunk(hunks[i], w_{st}, w_{end})$\;
        }

        $context \gets \DiffUtils.mergeHunks(newHunks)$\;
        \Return{$context$}
\caption{Adaptive Context-Expanding}\label{alg: context expanding}
\end{algorithm}

Algorithm~\ref{alg: context expanding} shows the details of adaptive context-expanding.
In Line 1, \detector{} defines the order of each candidate context width.
If two context width has the same priority, we choose the former one in this order.
In Lines 2-7, \detector{} defines the ``completeness'' of a given line number as the minimal depth of nodes (in AST) whose start line number equals the given line number.
Our rationale is that a context with a minimal start and end depth breaks the minimal number of statement blocks.
In Line 8, \detector{} invokes the method in \CodeIn{DiffUtils} to generate the change hunks when comparing the input files.
In Lines 9-12, it chooses the context width at the start point and end point of each changed hunk based on the preceding ``completeness'' method, \CodeIn{minDepth}.
In Line 13, \detector{} modifies the source code of \CodeIn{DiffUtils} to merge these code hunks into an integrated change file.

\subsection{Iterative Identification with Human Validation Results}
To ensure effectiveness, \detector{} adopts CodeBert~\cite{codebert} and StarCoder~\cite{starcoder} with 15.5B parameters to effectively generate the probability of each commit's fixing a vulnerability.
Here we choose CodeBERT  StarCoder as it is an open-source model and one of the SOTA Large Language Models for Code (Code LLMs).
Now that StarCoder is designed for generation tasks, we add a feed-forward neural network (FNN) layer after its original transformer layers.

\subsubsection{Workflow of \detector{}'s Model}
The input of this step consists of the commit message extracted from the given code commit and the code changes generated in the preceding step.
To avoid extremely large code commits, \detector{} truncates both the commit message and code changes of the \CodeIn{n-th} commit as follows:
\begin{equation}
    \begin{aligned}
        input[n] &=  commit\_messages[n][:max\_message\_size]\\
            &\ + code\_changes[n][:max\_code\_size]
    \end{aligned}
\end{equation}

Then, the propagation equation of each component in our StarCoder-based model is formally defined as:
\begin{equation}
    \begin{aligned}
        &h[n, idx] =  StarCoder(input[n])[,idx,]\\
        &x_n = FNN(\mathop{Average}\limits_{idx} (h[n, idx])) \\
        &\hat{x_n} = {1} / {(1 + exp(x_n)) } \\
    \end{aligned}
\end{equation}
where $h[n, idx]$ is StarCoder's last hidden state on the $idx-th$ token, $x_n$ is the prediction result after FNN, and $\hat{x_n}$ is the normalized probability of whether commit n fixes a vulnerability.
 
We employ weighted Focal Loss~\cite{lin2017focal} as our loss function.
It is calculated based on the prediction result $x$, label $y$, and two optional weight $\alpha$ and $\gamma$:

\begin{equation}~\label{eq: loss}
l_n = 
\begin{cases}
    - \alpha \times ( {(1 - \hat{x_n})}^{\gamma} \log \hat{x_n}) & y_n = 1\\
    - (1 - \alpha) \times ( {\hat{x_n}}^{\gamma} \log (1 - \hat{x_n})) & y_n = 0\\
\end{cases}
\end{equation}
 
where $y_n = 1$ indicates that the t-th library is affected by the input vulnerability, and otherwise $y_n = 0$.
The optional weight $\alpha$ is set as $0.9$ to amplify the loss of positive samples due to the imbalance between positive samples and negative samples.
The optional weight $\gamma$ is set as $4$ to amplify the loss of ``difficult'' samples to help \detector{} learn on these ``difficult'' samples.

\subsubsection{Selection Criteria for Human Validation}~\label{sec: human validation criteria}
Among the identified commits, i.e., marked as vulnerability patches by \detector{}, we require our security engineers to validate them with two criteria.
First, the validated code commits has a maximum number as human efforts is expensive.
Second, our security engineers validate these identified commits based on the descending order of \detector{}'s output probabilities, and they will stop validation and start another round of iterative fine-tuning when the precision is less than a given threshold, e.g., 0.9 (where the portion of false positives is larger than 10\%).

\section{Evaluation}~\label{sec: evaluation}
Our evaluation answers the following five research questions about \detector{}:
\begin{itemize}

    \item \textbf{RQ1}: How effectively does \detector{} identify Java vulnerability patches? 
    \item \textbf{RQ2}: What is the improvement of our iterative identification framework within the same human efforts?
    \item \textbf{RQ3}: What is the contribution of a comprehensive context to the effectiveness of \detector{}?
    \item \textbf{RQ4}: How efficient is \detector{} in identifying Java vulnerability patches?
    \item \textbf{RQ5}: What is the usefulness of \detector{} in real-world security practice?
\end{itemize}


\subsection{Dataset}
To evaluate the effectiveness of identifying silent vulnerability patches, we use the Java dataset proposed by recent work~\cite{vulfixminer}, in short as \dataset{}.
Specifically, it collects 1,474 commits of vulnerability patches from 150 Java repositories.
Then it collects the remaining commits from these 150 Java repositories as non-vulnerability patches, including 839,682 commits.
Among the remaining commits, it randomly selects 126,100 of them as negative samples after data cleaning~\cite{midas}.
\dataset{} is split in a cross-project setting, indicating that the repositories in the training and validation set do not occur in the testing set.
We follow the same portion between positive and negative samples as our baselines~\cite{vulfixminer, midas, deepjit}, i.e., 1:10 in training/validation set and 1:400 in testing set.

\dataset{} includes only changed lines for each included code commit.
We clone the source code of each included repository as \detector{} takes all files before and after one code commit in its repository.
We update 15 repositories in \dataset{}, including non-Java repositories, unavailable ones, and repositories whose names have been changed (as of May 15, 2024).
Then, we remove all duplicated commits in \dataset{}.
After updating repositories and removing duplication, there are 1,359 vulnerability patches and 82,265 non-vulnerability patches whose statistics are listed in Table~\ref{tab: dataset distribution}.

\begin{table}[t]
\centering
\small
\caption{Statistics of Cleaned \dataset{}}
\label{tab: dataset distribution}
\begin{tabular}{llrrrrccr}
\toprule
\multicolumn{1}{l}{Dataset}        &  Label    & \#Commit & \#File   & \#Line   & \#Token & \multicolumn{1}{l}{\#Project} \\
\midrule
\multirow{2}{*}{Training}   & positive & 1,001   & 2.3   & 735.3  & 6827.9         & \multirow{2}{*}{103}         \\
                            & negative & 10,010  & 2.2  & 787.0    & 7548.4        &                              \\
\midrule
\multirow{2}{*}{Validation} & positive & 112    & 2.0    & 734.5  & 6436.6       & \multirow{2}{*}{102}         \\
                            & negative & 1,120   & 2.2   & 793.5  & 7567.1         &                              \\
\midrule
\multirow{2}{*}{Testing}    & positive & 246    & 2.8    & 1049.4 & 8895.3          & \multirow{2}{*}{24}          \\
                            & negative & 71,135  & 2.1 & 860.1  & 8076.8        &                              \\
\midrule
Total                       & -        & 83,624  & 2.1 & 849.4  & 7992.0       &    127                          \\
\bottomrule
\end{tabular}
Columns [\#File, $\dots$, \#Line] represent the average number of each commit.
\end{table}

\begin{table}[t]
\centering
\caption{Hyper-Parameters Used in Training StarCoder}
\label{tab: hyper parameter}
\begin{tabular}{cccccc}
\hline
Optimizer & \makecell[c]{DeepSpeed\\ Config} & \makecell[c]{Learning\\ Rate} & \makecell[c]{Lora\\ Alpha} & \makecell[c]{Lora\\ Dropout} & \makecell[c]{Weight\\ Decay}\\
\hline
 adamw & zero 3 & 1e-4 & 32 & 0.05 & 0.05 \\
\hline
\end{tabular}
\end{table}

\subsection{Metrics}
We evaluate the effectiveness of patch identification models from two perspectives of metrics: effectiveness and practicality.
We use the area under the curve (AUC)~\cite{auc} as a representative effectiveness metric because AUC is designed for imbalanced data (e.g., the number of vulnerability patches is substantially less than that of non-vulnerability patches).
AUC is widely used by approaches~\cite{midas, vulfixminer, colefunda} that identify vulnerability patches.
AUC measures a model's prediction performance for all discrimination thresholds ranging from 0 to 1.
AUC is defined as the area under the receiver operating characteristic (ROC) curve, and a ROC curve plots this model's true-positive rates against its false-positive rates at various discrimination thresholds.
An empirical study~\cite{romano2011using} shows that an identification model that achieves an AUC score $\geq 0.7$ ($0 \leq AUC \leq 1$) can be considered effective.

We use the F1 score and Matthews correlation coefficient (MCC)~\cite{mcc} as two practicality metrics.
These metrics reflect the ratio of true positives (vulnerability patches) against false positives and false negatives, so they can be used to estimate the manual costs of security engineers to validate and confirm \detector{}'s identification results.
These metrics are also used in vulnerability patch identification~\cite{graphspd} and other binary classification tasks~\cite{wu2023plms, narayan2021discriminative}.


\subsection{Baseline Approaches for Comparison}
We compare \detector{} with four state-of-the-art (SOTA) baselines:  VulFixMiner~\cite{vulfixminer}, DeepJIT~\cite{deepjit}, Sun et. al.~\cite{sunVulPatch}, and MiDas~\cite{midas}.
Note that \colefunda{}~\cite{colefunda} identifies only function-level vulnerability patches and GraphSPD~\cite{graphspd} uses specialized characteristics of C/C++, so we do not consider them as our baseline approaches. 
To compare \detector{} with existing graph-based approaches, we implement a graph-based baseline, which is designed based on RGCN~\cite{chen2019rgcn} (better than another two widely-used ones, GGNN~\cite{ggnn} and GCN~\cite{gcn}) and uses CodeBERT for node embedding.

\subsection{Evaluation Environments}
We perform all the evaluations in the environment running on the system of Ubuntu 18.04.
We use one Intel(R) Xeon(R) Gold 6248R@3.00GHz CPU, which contains 64 cores and 512GB memory.
We use 8 Tesla A100 PCIe GPUs with 40GB memory for model training and inference.
The hyper-parameters used for model training are listed in Table~\ref{tab: hyper parameter}.

\subsection{\textbf{RQ1}: How effectively does \detector{} identify Java vulnerability patches? }~\label{sec: rq1}

\subsubsection{Methodology}
We evaluate the effectiveness of \detector{} and baselines on \dataset{} from the perspective of the preceding metrics, AUC, F1 (including Precision and Recall), and MCC.
To fairly compare with baselines, we choose CodeBERT as our classification model (the same as VulFixMiner).
We further conduct ablation studies on \detector{}'s three main components of context generation, multi-granularity slicing, adaptive context-expanding, and input generation of our model.

\subsubsection{General results}
Table~\ref{tab: baseline comparison} shows the scores of all evaluation metrics.
We show that \detector{} achieves substantially high performance in each metric.
Specifically, \detector{} achieves an AUC score of 0.944, which is a 
It also achieves an F1 score of 0.441 and an MCC score of 0.456, indicating that even in an extremely imbalanced dataset (the ratio of positive and negative samples equals 1: 300), \detector{} can still effectively identify vulnerability patches.

Compared to SOTA baselines, \detector{} achieves higher scores on all effectiveness metrics.
\detector{} improves the F1 score by 20\% (from 0.402 to 0.603).
Specifically, \detector{} achieves a substantially high precision score of 0.910 while SOTA baselines achieve at most 0.562.
These evaluation results illustrate that \detector{} has shown higher effectiveness in identifying vulnerability patches and such improvement is mainly because of utilizing human validation results.
As shown in Section~\ref{sec: human validation criteria}, we require our security engineers to validate only 250 identified results, so our achieved effectiveness arguably outweighs the small human efforts.


\begin{table}[t]
\centering
\caption{Effectiveness of \detector{} and Baselines}
\label{tab: baseline comparison}
\begin{tabular}{lccccc}
\toprule
Approach     & AUC   & Precision & Recall & F1 & MCC   \\
\midrule
VulFixMiner  & 0.807 & 0.362     & 0.283  & 0.317   & 0.318 \\
Sun et. al.  & 0.619 & 0.052     & 0.110  & 0.070   & 0.071 \\
DeepJIT      & 0.724 & 0.104     & 0.146  & 0.122   & 0.144 \\
MiDas        & 0.852 & 0.185     & 0.217  & 0.200   & 0.198 \\
RGCN         & 0.867 & 0.562     & 0.313  & 0.402   & 0.437 \\
\midrule
\detector{} & \textbf{0.874} & \textbf{0.910}     & \textbf{0.451}  & \textbf{0.603}   & \textbf{0.627}  \\
\bottomrule
\end{tabular}
\end{table}

\subsubsection{Ablation Studies }
Table~\ref{tab: ablation study} shows the results of our ablation study.
Overall, removing each component of context generation from \detector{} leads to a drop ($\downarrow$) in most metrics.
Specifically, the AUC drops by at least 3\% and the F1 score substantially drops by at least 12\%.
This result indicates that each component of our proposed context is necessary and contributes to the effectiveness of \detector{}.

\begin{table}[t]
\centering
\caption{Ablation Studies of Each Component in \detector{}'s Context Generation}
\label{tab: ablation study}
\begin{tabular}{lccccc}
\toprule
Ablation Studies & AUC   & Precision & Recall & F1    & MCC   \\
\midrule
{ w.o. slicing}        & 3\% $\downarrow$ &  3\% $\uparrow$   & 9\% $\downarrow$ & 5\% $\downarrow$ & 5\% $\downarrow$\\
{ w.o. context}        & 3\% $\downarrow$ & 12\% $\downarrow$   & 5\% $\downarrow$ & 9\% $\downarrow$ & 9\% $\downarrow$\\
{ w.o. message} & 5\% $\downarrow$ & 22\% $\downarrow$    & 14\% $\downarrow$ & 19\% $\downarrow$ & 18\% $\downarrow$\\
\bottomrule
\end{tabular}
\end{table}



\subsection{\textbf{RQ2}: What is the improvement of our iterative identification framework within the same human efforts?}~\label{rq: iteration}


\subsubsection{Methodology}
This research question aims to evaluate the effectiveness of our iterative identification framework.
We choose CodeBert (used by VulFixMiner) and StarCoder (used by \detector{}) for evaluation.
For each model, we conduct one to five rounds of iteration for evaluation.
For a specific number of iteration rounds, we set our evaluation metric as the number of identified vulnerability patches within the same human efforts, i.e., identifying 250 code commits in total, which is close to the number of vulnerability patches in the testing set, 246 ones.
As discussed in Section~\ref{sec: human validation criteria}, we conduct another round of fine-tuning based on the human validation results of identified results when the precision of identified results is less than the given threshold (0.9).
Specifically, we take the ground-truth labels of vulnerability patches as the human validation results.



\subsubsection{General Results}
\begin{figure*}[t]
\begin{minipage}{0.48\textwidth}
    \centering
    \includegraphics[width=1\linewidth]{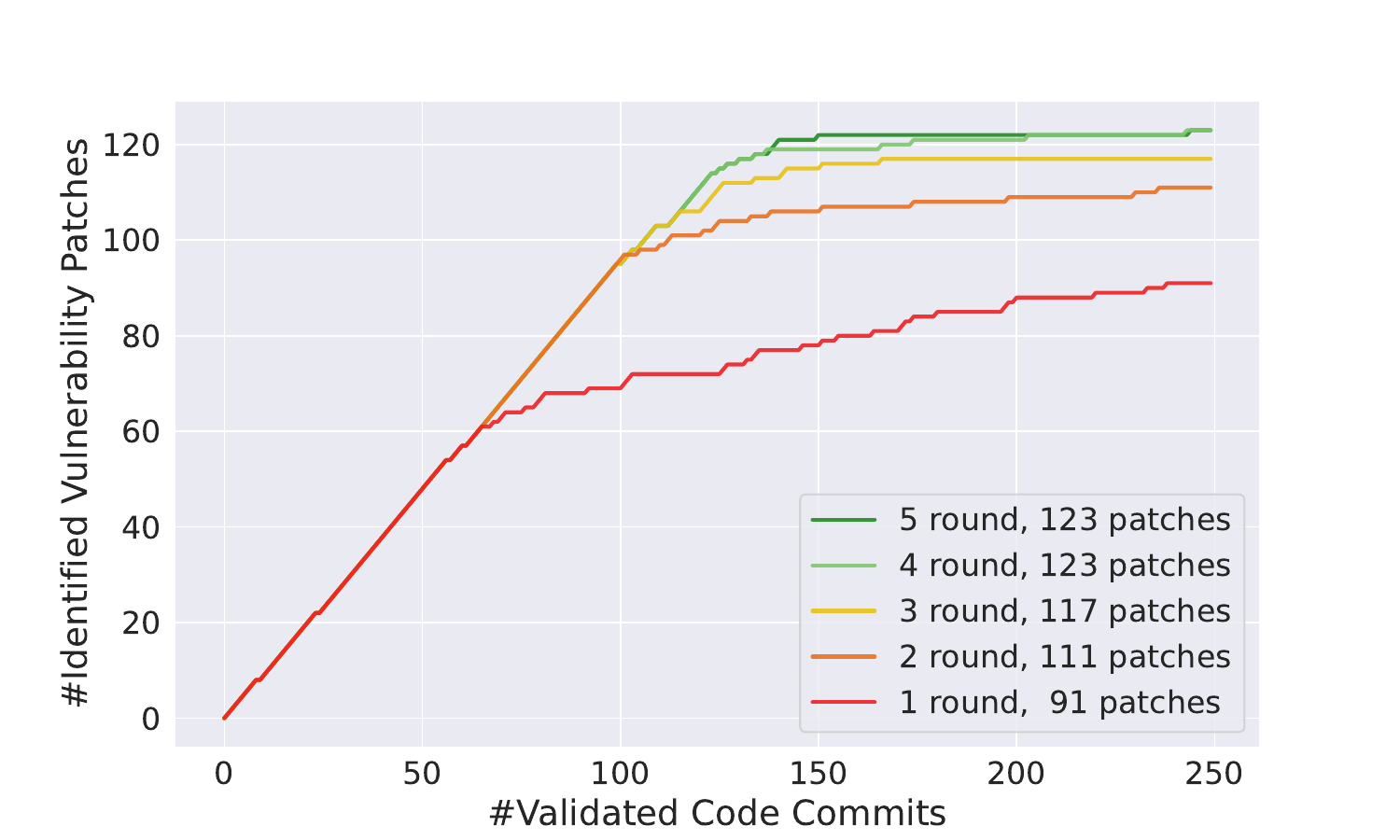}
    \captionsetup{justification = centerlast}
    \subcaption{CodeBERT's Effectiveness}
    \label{fig: iterative codebert}
\end{minipage}
\begin{minipage}{0.48\textwidth}
    \centering
    \includegraphics[width=1.\linewidth]{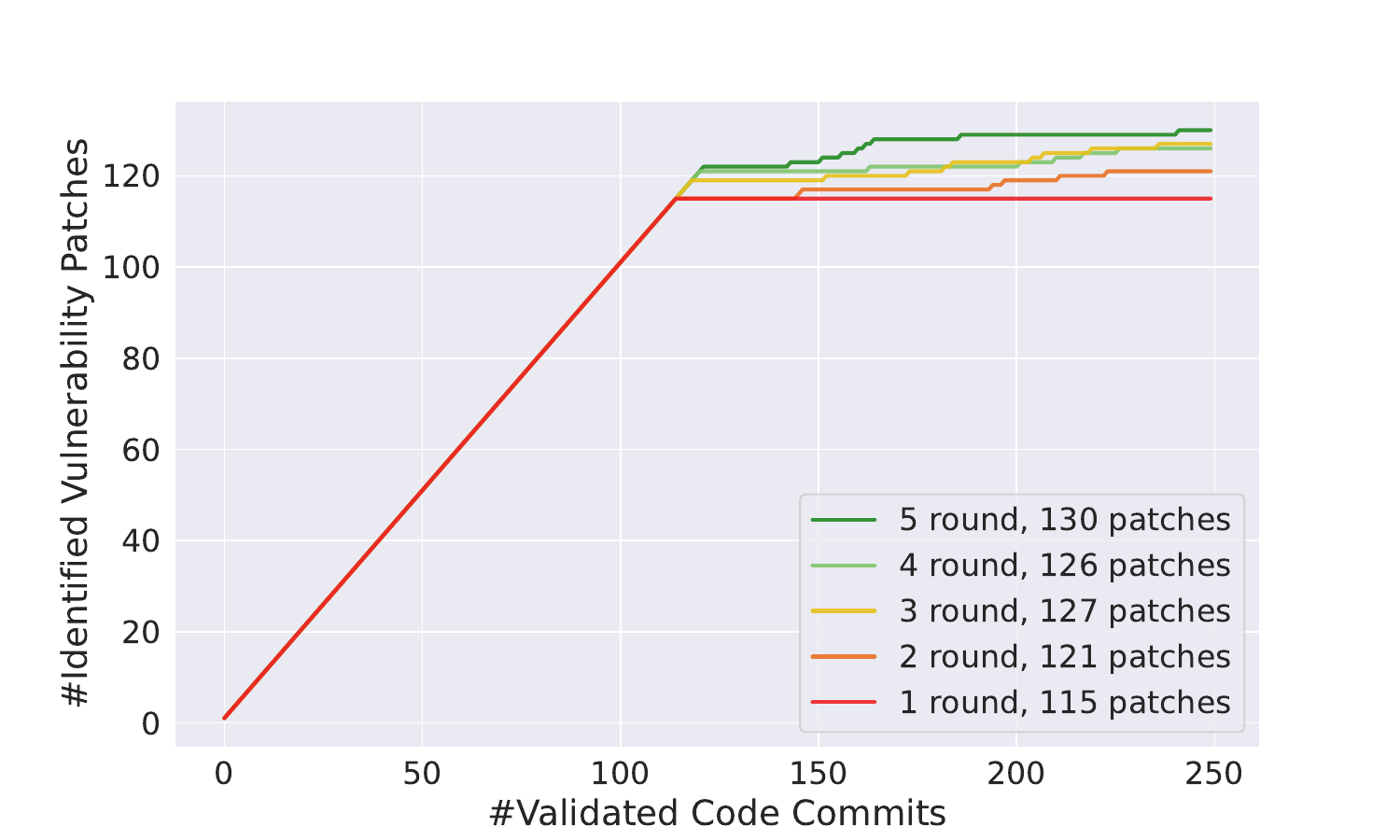}
    \captionsetup{justification = centerlast}
    \subcaption{StarCoder's Effectiveness}
    \label{fig: iterative starcoder}
\end{minipage}
\caption{Effectiveness of Iterative Identification Based on Human Validation Results}
\label{fig: iterative models}
\end{figure*}

Figure~\ref{fig: iterative models} shows the number of identified vulnerability patches with manual validation.
Compared to a single round of iteration (the red line), a five-round iteration (the green line) identifies 35\% more vulnerability patches with CodeBERT (as shown in Figure~\ref{fig: iterative codebert}) and 13\% more vulnerability patches with StarCoder (as shown in Figure~\ref{fig: iterative starcoder}).
This substantial improvement indicates that with the same human efforts, our iterative identification framework substantially improves the number of identified vulnerability patches, thus bringing high values in industrial practice.

\subsubsection{The Left Half of Each Curve}
The left half of each curve in Figure~\ref{fig: iterative models} displays a nearly constant slope, resembling a straight line. This indicates high precision during manual validation, thereby significantly reducing the costs of human efforts. Furthermore, as the number of iteration rounds increases, the length of this straight-line segment also extends. This suggests that our iterative framework maintains high precision, especially when our human efforts are limited to a small number of commits.


\subsubsection{Difference between CodeBERT and StarCoder}
We also show that our iteration framework improves more on CodeBERT than StarCoder, especially that from one round to two rounds.
The main reason is that compared to StarCoder, CodeBERT has less domain knowledge about open-source repositories, thus resulting a low F1 score in a cross-project setting.
However, \detector{} converts this cross-project setting to a mixed-project one by another round of fine-tuning based on human validation results, thus addressing this limitation.

\subsection{\textbf{RQ3}: To what extent does the context size of code commits affect the effectiveness of \detector{}?}~\label{rq: context}
Now that \detector{}'s contexts includes a hyper-parameter, the maximum context width and it highly depends on the window size of used LLMs, we also conduct evaluations on the combinations of these two hyper-parameters.

\subsubsection{Methodology}

In this research question, we choose StarCoder as our classification model.
For window sizes, we select three representative ones, 512, 1,024, and 2,048.
As for context widths, we also select three maximum widths, 3, 5, and 10 lines.
Then we evaluate the effectiveness of \detector{} with each combination of window sizes and context widths on \dataset{} from the perspective of the preceding metrics.
In the following of this paper, we denote \detectorWindowContext{size}{width} as \detector{} with the window $size$ of StarCoder and the maximum context $width$ of code commits.

We also evaluate the distribution of token numbers when the window size varies to directly show how each code commit with comprehensive contexts is truncated, thus affecting \detector{}'s effectiveness.
esides the preceding three maximum context widths, we also list the distribution of two constant context widths, 0 and 3, which are used in the baseline approaches~\cite{vulfixminer, midas} and \CodeIn{git diff}~\cite{git}, respectively.

\begin{table}[t]
\centering
\caption{The Effectiveness of \detector{} with Various Window Sizes and Context Widths in One Round of Iteration}
\label{tab: size and width}
\small
\begin{tabular}{ccccccc}
\toprule
\multicolumn{1}{l}{\makecell[c]{Window\\ Size}} & \makecell[c]{Max Context\\ Width} & AUC   & Precision & Recall & F1    & MCC   \\
\midrule
\multirow{3}{*}{512}            & 3                 & \textbf{0.925} & 0.359     & \textbf{0.309}  & \textbf{0.332} & \textbf{0.328} \\
                                & 5                 & 0.917 & \textbf{0.365}     & 0.297  & 0.327 & 0.327 \\
                                & 10                & 0.904 & 0.356     & 0.297  & 0.324 & 0.322 \\
\midrule
\multirow{3}{*}{1,024}           & 3                 & 0.914 & 0.299     & 0.280   & 0.289 & 0.287 \\
                                & 5                 & 0.917 & \textbf{0.462}     & 0.370   & \textbf{0.411} & \textbf{0.420}  \\
                                & 10                & \textbf{0.941} & {0.395}     & \textbf{0.390}   & 0.393 & 0.387 \\
\midrule
\multirow{3}{*}{2,048}           & 3                 & 0.879 & 0.212     & 0.220   & 0.216 & 0.215 \\
                                & 5                 & \textbf{0.944} & \textbf{0.528}     & 0.378  & 0.441 & \textbf{0.456} \\
                                & 10                & 0.928 & 0.470      & \textbf{0.419}  & \textbf{0.443} & 0.445 \\
\bottomrule
\end{tabular}
\vspace{-0.3cm}
\end{table}

\subsubsection{General results}
Table~\ref{tab: size and width} shows the effectiveness of \detector{} with various context widths and window sizes in one round of iteration.
We show that the context width of 3 is suitable for the window size 512 and the context width of 5 and 10 are proper for the window size of 1,024 and 2,048.

\begin{figure}[t]
\centering
\includegraphics[width=1\linewidth]{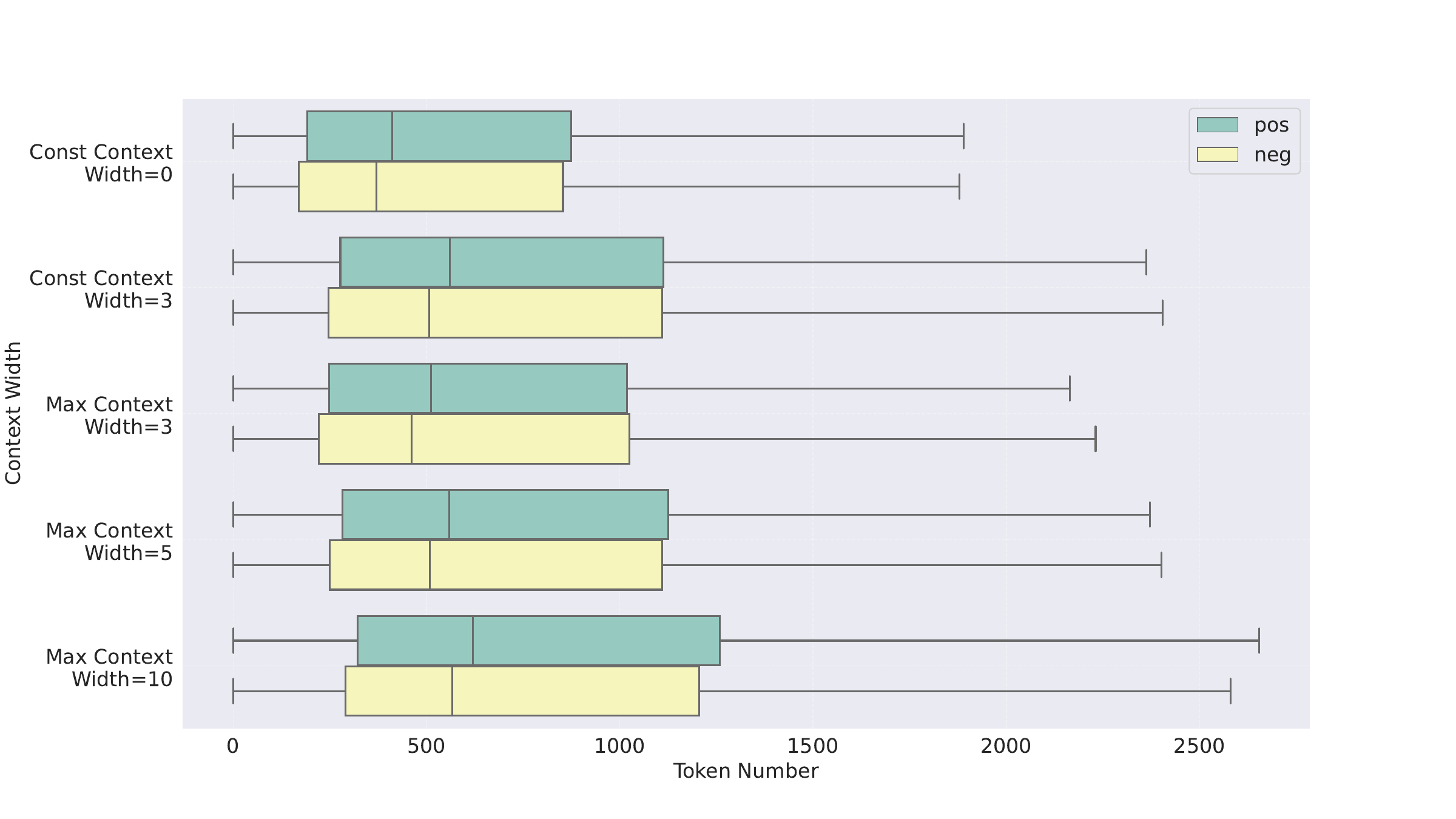}
\caption{The distributions of token numbers}
\label{fig: token}
\end{figure}

From this table, we show that \detectorContext{3} achieves higher scores in all metrics compared to \detector{} with other context widths.
For example, \detectorContext{3} achieves the F1 of 0.332 while the best AUC of other context widths is 0.327.
This difference can be explained by the distribution of token numbers in Figure~\ref{fig: token}.
This figure illustrates that the median token number in \dataset{} without contexts is close to the window size of 512.
Thus, including more contexts leads to more truncation, which decreases the effectiveness of \detector{} instead.

Table~\ref{tab: size and width} also shows that when the window size increases to 1,024 and 2,048, \detectorContext{5} and \detectorContext{10} achieve higher scores on all metrics.
Specifically, the F1 score of \detectorContext{5} increases from 0.327 to 0.411 and 0.443, and the F1 score of \detectorContext{10} increases from 0.324 to 0.393 and 0.443.
This improvement comes mainly from the increase in the window size, which can contain a large portion of contexts without truncation.
Figure~\ref{fig: token} also shows that when the window size equals 512, more than half of the inputs in the preceding two scenarios are truncated.
In contrast, when the window size increases to 1,024 or 2,048, only a small portion of inputs are truncated.
Thus, a window size of 2,048 is sufficient for our LLM to comprehend their semantics completely, resulting in higher identification effectiveness.

From Table~\ref{tab: size and width}, we also show that the effectiveness of \detectorContext{5} and \detectorContext{10} are quite similar under various window sizes.
For example, their maximum difference in F1 score is 1.8\% (0.411 - 0.393).
The main reason is that the token number increases to only a small extent when the context increases from 5 to 10.
From Figure~\ref{fig: token}, we show that their distributions are quite similar.
This result illustrates that most code changes have been expanded with complete contexts when the maximum context width is large enough, e.g., $\geq 5$.
Thus, a maximum context of 5 is sufficient for \detector{} to expand the contexts for code changes.

\subsection{\textbf{RQ4}: How efficient is \detector{} in identifying Java vulnerability patches?}~\label{sec: eval efficiency}
This research question evaluates the runtime overhead of \detector{} on \dataset{} with various window sizes, which is the key factor of training time.
For each window size, we select the maximum batch size so as to keep the same resource consumption (GPU memory).
We measure the end-to-end time of \detector{} with 10 epochs of training until its training losses converge. 

\subsubsection{General results}
The evaluation result is shown in Table~\ref{tab: efficiency}.
The end-to-end time consumption for \detector{} to identify a
given vulnerability is less than a half second.
Specifically, even when we use the largest window size of 2,048, \detector{} takes only 373.99 milliseconds on average to determine whether one code commit fixes a vulnerability.
As for the training consumption, the maximum training time of \detector{} is 65.30 hours, which is practical for real-world security practice.
Additionally, we also find that when we increase the window size to 8,000 it might cost more than 1,000 hours, which is unacceptable.
In general, \detector{} is efficient enough to identify Java vulnerability patches.


\begin{table}[t]
\centering
\caption{Time Consumption \detector{}}
\label{tab: efficiency}
\begin{tabular}{lccc}
\toprule
Model & \multicolumn{1}{c}{\makecell[c]{Training \\ (Hour)}} & \makecell[c]{Inference\\ (Hour)} & \multicolumn{1}{c}{\makecell[c]{Inference (ms)\\ per commit}} \\
\midrule
$\mbox{\detector{}}_{512}$  & 8.85 & 0.99   & 50.06         \\
$\mbox{\detector{}}_{1024}$ & 17.65   & 1.97     & 99.55  \\
$\mbox{\detector{}}_{2048}$ & 65.30     & 7.42    & 373.99     \\
\bottomrule
\end{tabular}
\end{table}

\subsection{\textbf{RQ5}: What is the usefulness of \detector{} in real-world security practice?}~\label{sec: eval practice}
\subsubsection{Methodology}

To evaluate \detector{}'s impact on security practice, we use \detector{} to scan new code commits of 5 widely used Java repositories (namely \CodeIn{jenkins, hutool, dubbo, hiro}, and \CodeIn{light-4j}).
For each repository, we scan the recent 500 code commits before July 2023.
Then, we manually inspect each identified code commit to determine whether it fixes a vulnerability.

\begin{figure}[t]
    \centering
 \begin{lstlisting}[style=base, caption = An example Vulnerability Patch of SECURITY-2399 (Jenkins), label = lst: security-2399]
!--- spring-core/src/main/java/hudson/search/Search.java!
~+++ spring-core/src/main/java/hudson/search/Search.java~
@@ -139,12 +144,15 @@
   public SearchResult getSuggestions(StaplerRequest req, String query) {
     SearchResultImpl r = new SearchResultImpl();
!-    int max = req.hasParameter("max") ? Integer.parseInt(req.getParameter("max")) : 100;!
~+    int max = Math.min(~
~+      req.hasParameter("max") ? Integer.parseInt(req.getParameter("max")) : 100,~
~+      MAX_SEARCH_SIZE~
~+    );~
     SearchableModelObject smo = findClosestSearchableModelObject(req);
     for (SuggestedItem i : suggest(makeSuggestIndex(req), query, smo)) {
       if (r.size() >= max) {
         r.hasMoreResults = true;
         break;
       }
       if (paths.add(i.getPath()))
         r.add(i);
     }
`$\dots$`
\end{lstlisting}
\vspace{-0.3cm}
\end{figure}


  

\subsubsection{General Results}
Generally, \detector{} identifies 54 candidate commits.
According to our manual inspection, 20 of them are vulnerability patches, 18 of them are fixes of high-risk bugs, and 10 of them are irrelevant commits (e.g., improvement).
The results can be found on our anonymous website~\footnote{\href{https://anonymous.4open.science/r/patchidentification}{https://anonymous.4open.science/r/patchidentification}~\label{repo url}}.

Listing~\ref{lst: security-2399} shows an example vulnerability patch~\footnote{\href{https://github.com/jenkinsci/jenkins/commit/d034626fd5290565f130f98adccd7ca519a2d10d}{https://github.com/jenkinsci/jenkins/commit/d034626fd5290565f130f98adccd7ca519a2d10d}} without a CVE ID.
In this example, its code changes add an upper bound (\CodeIn{100}) for the maximum length, \CodeIn{max}.
Additionally, its following context indicates that it is given by input requests and set as the maximum length of response suggestions.
With the help of these contexts, \detector{} identifies this code commit as one patch of an Out-Of-Memory (OOM) vulnerability.

\section{Discussion}~\label{sec: discuss}
\textbf{Distinguishing vulnerability patches from bug fixes.}
Now that vulnerabilities and bugs are highly overlapped~\cite{canfora2020investigating, camilo2015bugs}, vulnerability patches might be hard to distinguish from bug fixes, especially fixes of high-risk bugs.
Additionally, these high-risk bugs might be exploited as part of the vulnerabilities in the future, so these commits of bug fixing also help developers avoid potential risks.
Thus, we report 16 fixes of high-risk bugs identified by \detector{} separately in Section~\ref{sec: eval practice}. 

We provide an example fix of a high-risk bug~\footnote{\url{https://github.com/jenkinsci/jenkins/commit/3fe425f29dc67863d95fdfbefaf65cbc12391853}} identified by \detector{}.
As listed in Listing~\ref{lst: bug fix}, this commit removes the \CodeIn{getPassword()} method from its signature.
It seems to fix a vulnerability of password leakage, but it cannot be utilized by malicious users.
Thus, it is classified as a bug fix.

\begin{figure}
    \centering
 \begin{lstlisting}[style=base, caption = An Example Non-Vulnerability Patch from Jenkins, label = lst: bug fix]
!--- core/src/main/java/hudson/security/TokenBasedRememberMeServices2.java!
~+++ core/src/main/java/hudson/security/TokenBasedRememberMeServices2.java~
@@ -204,7 +201,6 @@ 
     UserDetails userDetails = getUserDetailsService().loadUserByUsername(cookieTokens[0]);
!-    String expectedTokenSignature = makeTokenSignature(tokenExpiryTime, userDetails.getUsername(), userDetails.getPassword());!
~+    String expectedTokenSignature = makeTokenSignature(tokenExpiryTime, userDetails.getUsername());~
     if (`!`equals(expectedTokenSignature, cookieTokens[2])) {
       throw new InvalidCookieException("Cookie token[2] contained signature '" + cookieTokens[2]
       + "' but expected '" + expectedTokenSignature + "'");
     }
\end{lstlisting}

\end{figure}

\section{Threats to Validity} \label{sec:threats}

A major threat to external validity comes from the labels of \dataset{}.
In Section~\ref{rq: iteration}, we take the ground-truths of each commit as human validation results.
However, in real-world practice, security engineers might face difficulties or incorrectly validate the identified commits, thus decreasing the end-to-end effectiveness of \detector{}.
To mitigate this threat, in Section~\ref{sec: eval practice}, we choose five senior security engineers with at least two years experience for validation.
Thus, their validation results can be representative as ground-truths of our identified commits.
The threats to internal validity are instrumentation effects that can bias our results.
To reduce these threats, we manually inspect the intermediate results such as the comprehensive contexts for dozens of sampled vulnerability patches, such as those in Listing~\ref{lst: cve-2017-4971-reduce} and ~\ref{lst: security-2399}.

\vspace{-0.2cm}
\section{Related Work}
\subsection{Vulnerability Patch Identification}
Patch Identification~\cite{vulfixminer, vulcurator, espi, zhou2021spi, colefunda, graphspd} identifies whether a code commit corresponds to a vulnerability fix.
It helps users to be aware of vulnerability fixes and apply fixes in time because a vulnerability in open source software (OSS) is suggested to be fixed ``silently'' until the vulnerability is disclosed. 
VulFixMiner~\cite{vulfixminer}, is a representative approach of identifying Java and Python vulnerability patches.
Given a code commit, VulFixMiner extracts the removed and added codes, and uses a pre-trained model, CodeBert~\cite{codebert}, to encode them.
Then, it uses a fully connected layer to determine whether this code commit corresponds to a vulnerability fix.
However, these existing approaches do not consider comprehensive contexts of code commits, thus suffering from low accuracy.
Thus, we design \detector{} with comprehensive contexts of commit-level code changes to address this limitation.

\subsection{Vulnerability Mining}
Vulnerability Mining~\cite{deepvd, reveal, devign, ample, funded} identifies whether a code commit corresponds to a vulnerability fix.
It directly helps users to be aware of the vulnerability itself, which is the most emergent to be fixed.
Thus, many efforts have been spent to design vulnerability mining approaches, including fuzzing approaches~\cite{yu2022htfuzz, you2017semfuzz, chen2018iotfuzzer, hu2021achyb}, clone-detection approaches~\cite{kim2017vuddy, xiao2020mvp} and DL approaches~\cite{deepvd, reveal, funded, ample, devign, mvd, linevul}.
Devign~\cite{devign} is a representative one of DL approaches.
It uses Code Property Graphs to represent an input method and designs a specialized graph model to determine whether this method is vulnerable.
Most of these approaches focus on method-level/line-level vulnerability mining, with an assumption that all changed functions before a vulnerability patch are vulnerable.
However, this assumption is unreasonable because a patch commit can also include irrelevant changes~\cite{midas}.
Thus, they need further vulnerability localization to improve data quality.

\subsection{Large Language Models}
Large Language Models (LLMs) mainly follow the Transformer~\cite{vaswani2017attention} structure with an encoder to encode a given input and a decoder to generate output sequences from the encoded representation.
These LLMs include billions of parameters and are pre-trained on billions of language corpus online.
In recent years, LLMs, such as GPT-4~\cite{openai2023gpt4}, LLaMa~\cite{llama}, and StarCoder~\cite{starcoder}, have been widely applied in software engineering~\cite{chen2021evaluating, xia2022practical} and security~\cite{chan2023transformer, tony2023llmseceval} tasks.
Additionally, LLMs can comprehend the semantics of code changes' raw text, thus improving the effectiveness of downstream tasks, such as vulnerability identification~\cite{chen2023diversevul, jain2023code}, vulnerability type classification~\cite{liu2023not}, and code review~\cite{li2022codereviewer}.

\section{conclusion}\label{sec:conclusion}
In this paper, we have proposed two novel algorithms to generate precise contexts for code commits, enhancing the accuracy of vulnerability patch identification.
Based on the preceding algorithms, we have designed \detector{}, an iterative identification framework utilizing human validation results.
We have conducted a comprehensive evaluation to demonstrate \detector{}'s effectiveness and efficiency for identifying vulnerability patches, improving the F1 score by 20\% compared to the best scores of the SOTA approaches.
We have demonstrated the effectiveness of our iterative identification framework, identifying 35\% (CodeBERT) and 13\% (StarCoder) more vulnerability patches within the same human efforts.
We have demonstrated  \detector{}'s high value to security practice by helping identify 20 vulnerability patches and 18 fixes of high-risk bugs from 2,500 recent code commits of five highly popular open-source projects.



\bibliographystyle{ACM-Reference-Format}
\bibliography{sample-base}

\end{document}